\documentclass[%
 reprint,
 amsmath,amssymb,
 aps,
]{revtex4-2}
\usepackage{svg}
\usepackage{graphicx}
\usepackage{dcolumn}
\usepackage{bm}
\usepackage{lipsum}

\begin{document}

\preprint{APS/123-QED}

\title{Strain Induced Kramers-Weyl Phase in III-V Zinc Blende Systems}

\author{Denis Aglagul\textsuperscript{1}}
\author{Jian Shi\textsuperscript{1,2}}%
 \email{Correspondence: shij4@rpi.edu}
\affiliation{\textsuperscript{1}Department of Physics, Applied Physics, and Astronomy, Rensselaer Polytechnic Institute, Troy, NY, 12180;\\
\textsuperscript{2}Department of Materials Science and Engineering, Rensselaer Polytechnic Institute, Troy, NY, 12180
}

\date{\today}

\begin{abstract}
We present theoretical observations on the topological nature of strained III-V semiconductors. 
By $k \cdot p$ perturbation, it can be shown that the strain-engineered conduction band hosts a Kramers-Weyl node at the $\Gamma$ point. It is theoretically shown a curated strain can create and then tune the sign of the topological charge. 
Furthermore, we outline experimental methods for both the realization and detection of strain-induced topological phase transitions.
\end{abstract}

\maketitle

\section{\label{sec:level1}Introduction}
Topological condensed matter systems highlight an intersection between fundamental physics and practical technologies. On the fundamental side we observe emergent quasi-particles such as Dirac or Weyl fermions; with them we can explore exotic physics such as topological charge transport and chiral anomaly \cite{armitage2018weyl}. The practicality manifests through motivations to increase the efficiency of modern electronics via engineering resistance-area product \cite{lanzillo2024topological}, low energy switching in spintronics 
 \cite{manipatruni2019scalable}, and the creation of impurity resistant devices. This further drives innovations in the design and engineering of topological quantum computing platforms such Majorana zero modes and fractal quantum hall states \cite{KITAEV20032,PhysRevLett.94.166802}. \\

Mirror symmetry can be broken by the application of curated strains. For example, in Fig. 1A,  a pristine crystal, marked in dashed lines, has two mirror symmetry elements \textit{m}$'$; by applying a uniaxial planar force along any directions that are not x, z, nor $<$101$>$, one can remove these two mirror symmetry elements. 
Such forces can be generalized to 3D as shown in Fig. 1B. In the example of cubic zinc blende (ZB), the crystal lacks inversion symmetry in its natural state, tuning the pictured $\theta$ and $\phi$ breaks the mirror and rotoinversion symmetries of the point group making the crystal chiral.

\begin{figure}
\includegraphics[width = .45\textwidth]{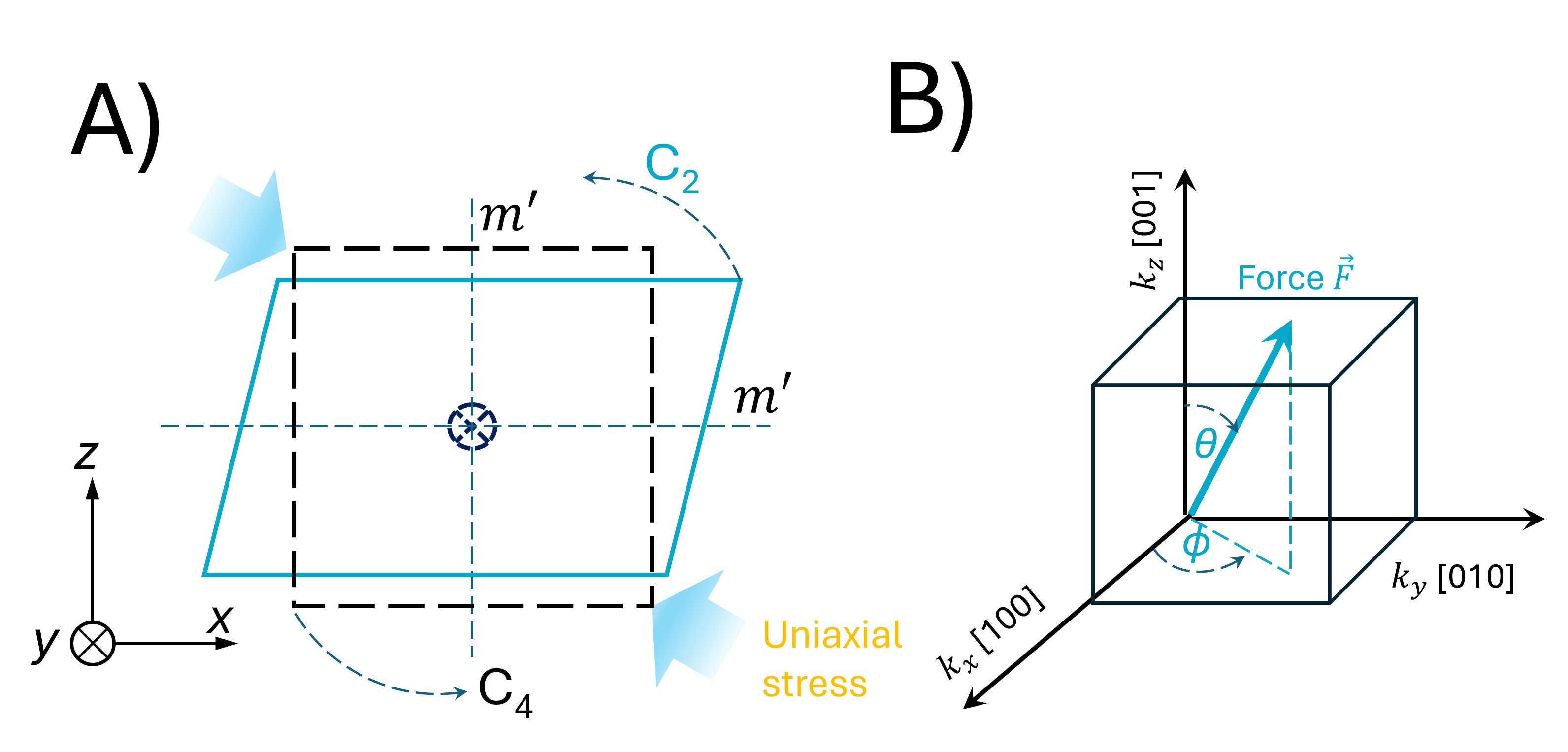}
\caption{\label{fig:intro} \textbf{A)} The removal of symmetry elements m$'$ due to a distortion in the crystal.
\textbf{B)} A 3-Dimensional force as the generator of the crystallographic distortion due to strain.
}
\end{figure}

\section{Zinc Blende as a case study}
A good case study is the cubic ZB crystal system which, as opposed to diamond, lacks an inversion center. The spin-orbit coupling in such bulk semiconductors can be written as a sum of the contributions of its inherent inversion-asymmetry \cite{dresselhaus1955spin} plus a strain induced component:
\cite{PhysRevB.16.2822,wu2010spin,bir1974symmetry,howlett1977effect,la1988effect}. 

\begin{equation}
H = H_0 + \Omega_{Dress} \cdot \sigma + \Omega_{Strain} \cdot \sigma
\label{eq:one}
\end{equation}

\begin{equation}
\Omega_{Dress}^x = \gamma \textit{k}_x(\textit{k}_y^2-\textit{k}_z^2)
\end{equation}

    \begin{equation}
          \Omega_{Strain}^x = C(\epsilon_{xy}\textit{k}_y - \epsilon_{xz}\textit{k}_z) + D\textit{k}_x(\epsilon_{zz} - \epsilon_{yy}) 
          \label{eq:strain}
    \end{equation}
where $H_0 = (\hbar \textit{k})^2/2m$, and $\epsilon_{ij}$ is the symmetric strain tensor.\\

In the above, $\gamma$, C, and D are material constants \cite{bernevig2005spin}. The prefactor $\gamma$, commonly called the Dresselhuas coefficient, reflects the cubic in \textit{k} spin-orbit coupling strength of ZB crystals. The influence of the second term $\Omega_{Strain}$ has been studied extensively by Seiler et al. via Shubnikov de Haas oscillations and showed both experimentally and numerically the influence of a few kilobar of pressure on the Fermi surface and conduction band-splitting in InSb \cite{seiler1977inversion}. Further discussion on the coefficients  C, and D can be found in \cite{la1988effect, bernevig2005spin,bernevig2006quantum,wu2010spin}. \\


\begin{figure*}

\includegraphics[width = .85\textwidth]{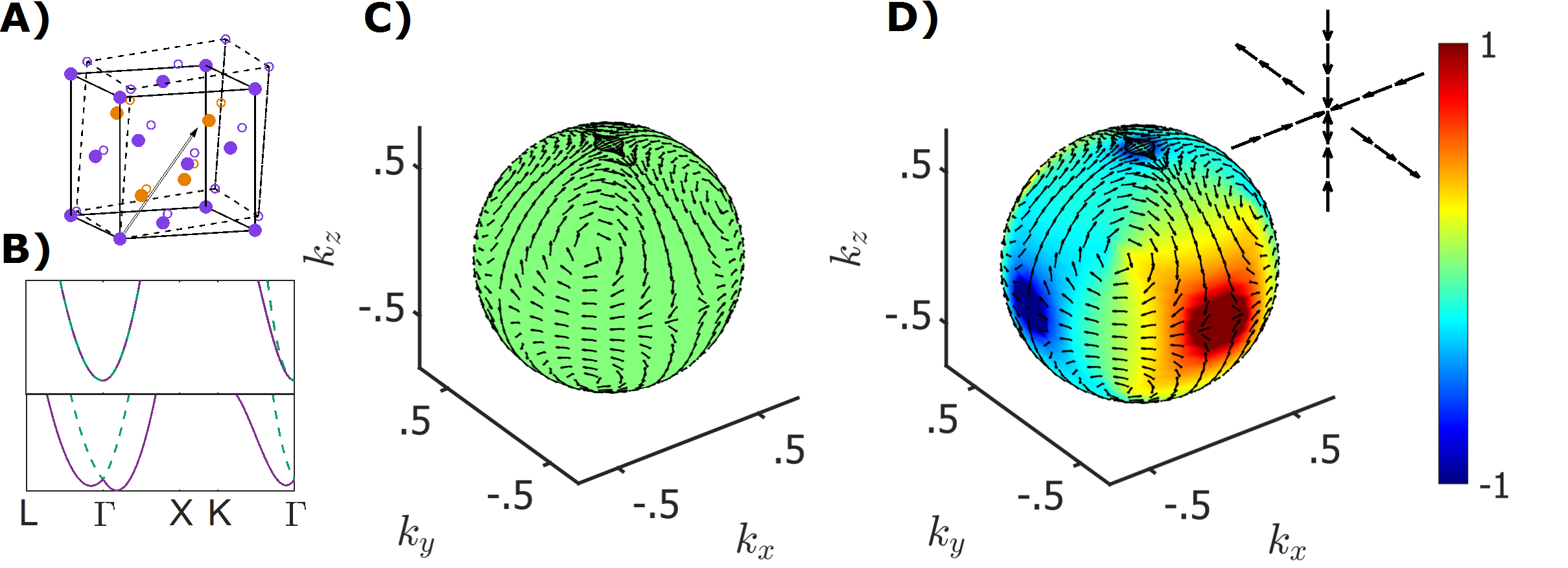}
\caption{\label{fig:spin}The influence of mirror/rotoinversion symmetry-breaking strain on band structure and spin-texture. \textbf{A)} A diagram showing the displacement of atoms in ZB when acted on by an exaggerated force. \textbf{B)} A general ZB band structure with spin degeneracy along $\Gamma - L$ and $\Gamma - X $ when mirror/rotoinversion symmetry is present, and spin-split when broken. The y axis is in the order of meV to eV. \textbf{C-D)} The spin texture painted on a uniform sphere with an arbitrary radius (in units of $m^{-1}$) in \textit{k}-space with care taken to respect any approximations necessitating small \textit{k} values. The spin texture on the \textit{k}-sphere is show as arrows and\textemdash for emphasis\textemdash the spin texture perpendicular to the surface is drawn as the colormap. Of the two, the left sphere is presented as an unstrained comparison and reflects precisely  Dresselhaus spin-orbit coupling behaviour. The rightmost sphere is the result of a force acting on a ZB crystal at $(\theta,\phi) = (\frac{\pi}{2},\frac{2\pi}{9})$. The coefficient C is suppressed and D = $\gamma$.}

\end{figure*}

In this letter we will examine the topological properties encoded within Eq. \ref{eq:strain}. Before discussing the more exotic effects of strain, we point out that any force will have an effect on the severity of the bulk inversion-asymmetry, elevating the corresponding spin-orbit coupling strengths. Fig. \ref{fig:spin} illustrates the action of strain on a ZB system by means of spin texture and band structure. Fig. \ref{fig:spin}A displays a cartoon of an exaggerated strain on the  crystal structure. In B), the presence of mirror/rotoinversion symmetry in the time reversal invariant ZB lattice is reflected by the spin degeneracy along $\Gamma-L$ and $\Gamma-X$ as seen in the band structure plots (top); the breaking of mirror/rotoinversion symmetry results in the spin splitting when tracing the plotted \textit{k}-vectors (bottom). The spheres in Fig. \ref{fig:spin} visualize the spin texture of the same two systems with the normalized colormap showing the spin components perpendicular to the surface. In the unstrained prototype, all spin vectors are parallel to this surface which remains true for a sphere of any radius centered at \textit{k} = 0. Dresselhaus spin orbit coupling does not contribute to the Chern number which is obtained by the Berry curvature flux through the Brillouin zone. Lastly, the strained spin texture is painted on the final sphere with the principal spin vectors presented as an inset\textemdash inspection of which reveals that our bounding sphere encloses a positive topological charge.  The spin-textures plotted in Fig. \ref{fig:spin} are for when C is suppressed. In a real material where C $>$ 0 and if shear strains are present, the principal spin textures start to have components that are parallel to their respective surfaces. It is observed that the normal strains along the three principal axes are the key for leading to the inward or outwards spin textures. As it is shown in the following section, we can engineer $\mathcal{C}_{Weyl}$= -1, +1 and trivial states by manipulating the strain.\\

Reference \cite{chang2018topological} showed that any non-magnetric chiral crystal harbors Kramers-Weyl nodes, coupled with knowledge that any crystal lacking an inversion center can always be strain engineered to be chiral \cite{peiser1963reduction}, we have a path forward to the realization of topological states by straining otherwise trivial semiconductors.\\

The Hamiltonian that describes a Weyl node is 

    \begin{equation}
        H_{Weyl} = A_{ij} k_i \sigma^j
        \label{eq:weyl}
    \end{equation}
which has similar structure as Eqs (\ref{eq:one},\ref{eq:strain}). Examining like terms, we can populate coefficients of a strain induced Weyl Hamiltonian. Explicitly,

\begin{equation}
A_{ij}^{strain} = \begin{pmatrix}
D(\epsilon_{zz} - \epsilon_{yy}) & -C\epsilon_{yx} & C\epsilon_{zx}\\
C\epsilon_{xy} & D(\epsilon_{xx} -\epsilon_{zz}) & -C\epsilon_{zy}\\
-C\epsilon_{xz}& C\epsilon_{yz} & D(\epsilon_{yy} -\epsilon_{xx})
\label{eq:A}
\end{pmatrix} 	    
\end{equation}
Notice that the Kramers-Weyl node is completely characterized by elements of the the strain tensor, C and D. Furthermore, the topological charge can be evaluated given just Eq. \ref{eq:A} via $\mathcal{C}_{Weyl} = SGN(DET(A)) 
 $\cite{armitage2018weyl,chang2018topological}. \\ 

\begin{figure}
    \centering
    \includegraphics[width = .45\textwidth]{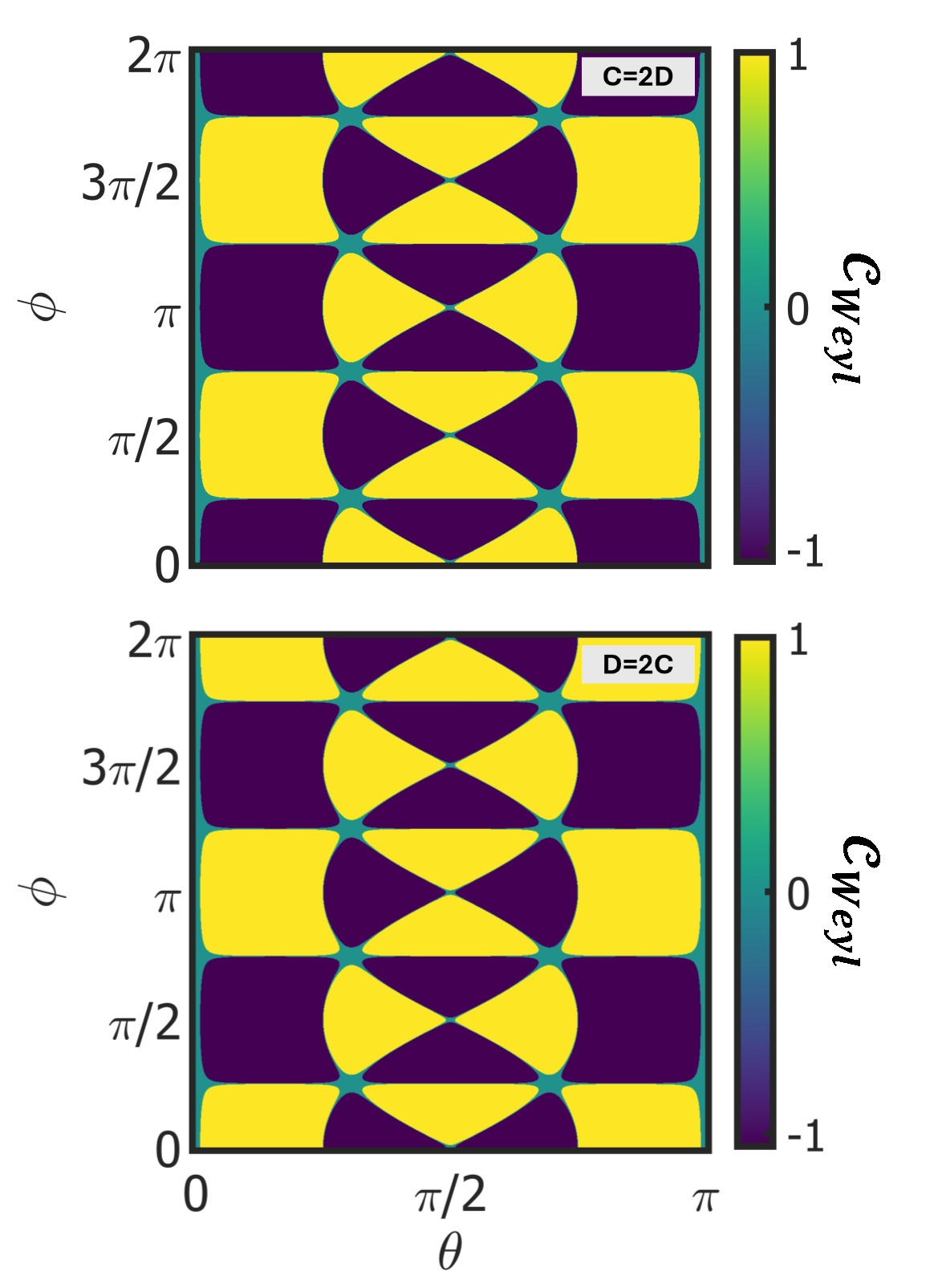}
    \caption{\label{fig:TPD}Topological phase diagram of a ZB  GaAs lattice being acted on by a force $\mathbf{F(\theta,\phi)}$. Top row is the case of C $=$  2D and the bottom row is D $=$ 2C. Small values are thresholded for numerical stability.}

\end{figure}

A uniaxial force along directions such as $<$100$>$, $<$110$>$, and $<$111$>$ will preserve some mirror/rotoinversion symmetries in the point group of the pristine lattice, thus we expect the determinant of Eq. \ref{eq:A} to vanish in these scenarios. Beyond these special test cases, it is not obvious which force vectors, $\mathbf{F(\theta,\phi)}$, transform the material into which topological state. As a means to explore this phase space, we create a topological phase diagram. By considering a single crystalline material system, the 6 degrees of freedom in the strain tensor are reduced to 2, the azimuthal and polar angles $0\le \phi \le 2\pi $, and $\le \theta \le \pi$ (assuming it is a tensile force), respectively. A set compliance tensor will translate $\mathbf{F(\theta,\phi)}$ into a symmetric strain tensor. The direction of this force, described in spherical coordinates, alone defines the topological state of the strained crystal\textemdash as seen in Fig. \ref{fig:intro}. For a practical force, we mean a reasonable pressure maybe on the order of $10^5$ to $10^6$ kPa being acted upon the crystal. Flipping the direction of the force flips the sign of the strain. The materials we investigate include GaAs and InSb due to their relative popularity and moderate Young's moduli, though the results still apply for any III-V ZB crystal. We evaluate SGN(DET(A)) for the range of angles mentioned and display the resulting topological phase diagram in Fig. \ref{fig:TPD}. Here we show the two main cases by plotting C = 2D, D = 2C in the top and bottom panel respectively. The transition between the top and bottom panel is not symmetric about unity for C/D. The exact point will depend on the compliance tensor. In physical situations it is seen that in GaAs the shear components dominate owing to C, the result is this material always behaves as the top panel. The compliance tensor used in Fig. \ref{fig:TPD}  belongs to GaAS \cite{de2015charting}. Practically, for GaAs, $\gamma$, C, D could be 8.5 $eV \AA^3$, 6.8 $eV \AA$, 2.1 $eV \AA$ respectively \cite{bernevig2005spin,wu2010spin,la1988effect}. \\

The nature of Fig. \ref{fig:TPD} can be interpreted by re-examining the previously discussed special cases. Forces lying within the {110} plane and principal axes do not create a chiral structure. The model remains topologically trivial for forces along $\phi = (2n+1)\frac{\pi}{4}$ for all $\theta$, $n \in \mathbb{Z}$ . Additionally, the Chern number vanishes for forces enacted along $sin(\phi) = \pm cot(\theta),cos(\phi) = \pm cot(\theta)$. These states must correspond to forces which permit some mirror/rotoinversion symmetries to exist in the deformed lattice.\\

\section{Experimental Outlook}

Strain engineering is a mature research field. In particular, epitaxial strain is a well-known method for the manipulation of material properties \cite{schlom2014elastic, li2014elastic}. Today, around 90\% of produced silicon transistors are strained to meet desired specifications \cite{bedell2014strain}. Piezoelectrics are another option if tunability is desired. Samples can either be mounted directly on top of piezoelectrics or be prepared in specialized devices \cite{hicks2014piezoelectric}. Regardless of the preferred method, there is also the issue of the maximal strain a bulk sample can sustain. Nanomembranes show a theoretical limit of around 10\% \cite{jiang2022strain} though bulk samples will be much lower due to the increased likelihood of defects. Whether or not a maximal strain is sufficient depends on the material properties.\\

\begin{figure}
    \centering
    \includegraphics[width = .45\textwidth]{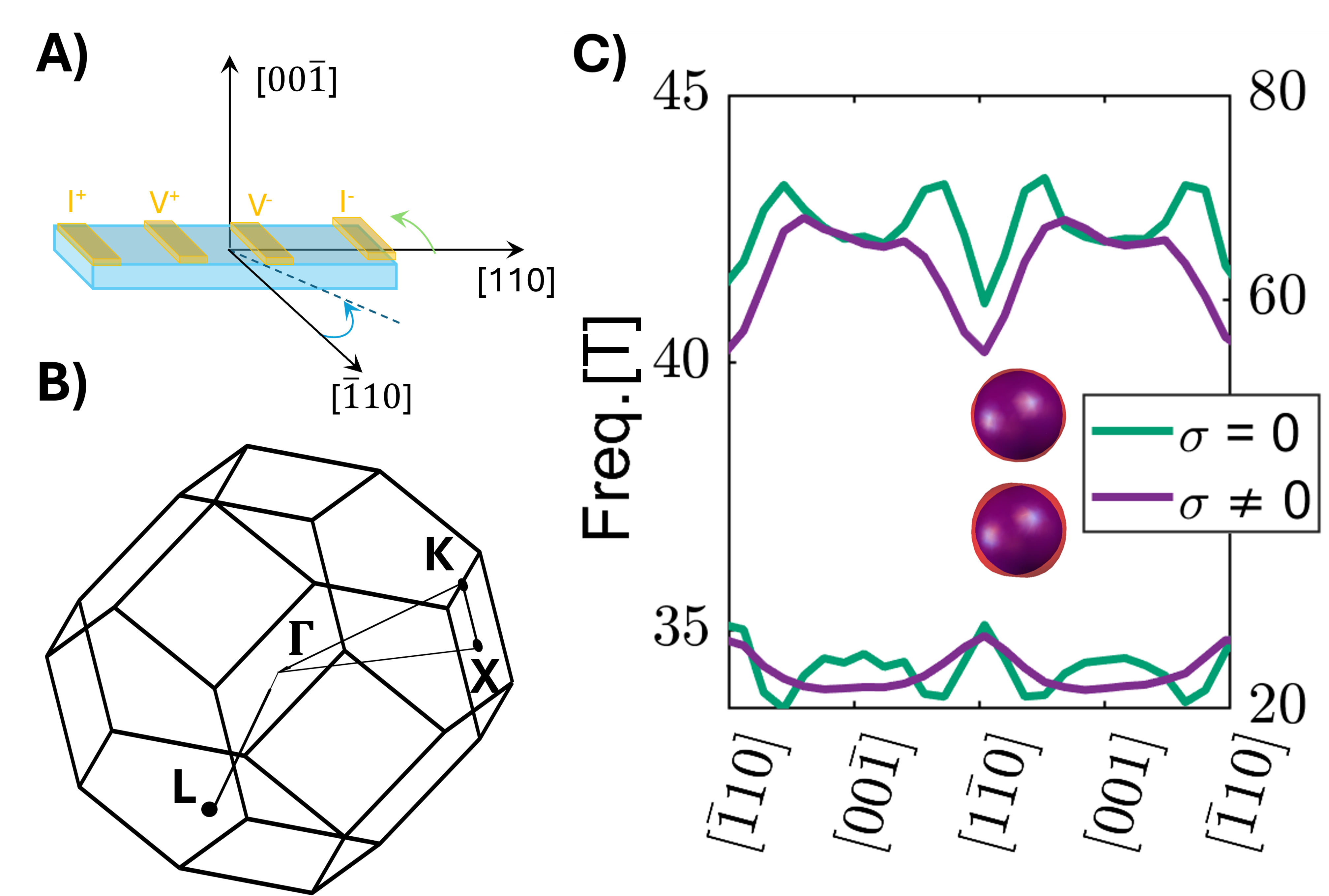}
    \caption{\label{fig:SdH}An experimental perspective. \textbf{A)}Diagram illustrating a 4-terminal angle resolved magneto-resistance measurement. The scanning angle is defined as between $[\bar110]$ direction and the magnetic field vector. \textbf{B)}First Brillouin Zone for zinc blende crystal structure with 4 symmetry points highlighted. \textbf{C)} A Monte Carlo simulation of quantum oscillation frequency vs angle for $
    \sigma = 0$ (green, left axis) and $\sigma \neq 0$ with strain on the order of 0.1\% and $\theta=85^o,\phi=15^o$ (purple, right axis).}
    
\end{figure}

A difficulty adjacent to growing/creating these test systems is the characterization of the desired topological properties. In the case of most III-V semiconductors there are some liberties that can be taken; since we have a 1:1:1 tri-correspondence between the existence of the Kramers-Weyl node, mirror/rotoinversion symmetry breaking, and chirality, the identification of any of these is equivalent to the detection of the others.
In the realm of the optical methods there is Angle Resolved Photoemission Spectroscopy \cite{sobota2021angle,xu2015discovery} for the detection of topological states, and  Circular Photo-Galvanic effect measurements for the detection of chirality and Weyl structures \cite{ma2017direct,rees2020helicity}. Other possible avenues which have had substantial attention recently include Faraday rotation and Magneto Optical Kerr Effect experiments, chiral anomaly, and bilinear magnetoresistance measurements. All of these have already had bountiful discussion in literature so we will not discuss them here \cite{bernevig2005spin,kato2004coherent,xiong2015evidence,calavalle2022gate,he2018bilinear}. Instead, we will explore the strain-tuned quantum oscillations in III-V semiconductors as a direct measurement of the cross-sectional area of the Fermi surface \cite{xiang2015angular,murakawa2013detection,moll2016transport,seiler1977inversion}. If we take the magnetic field direction as the \textit{z} axis, the oscillation frequency is proportional Fermi-surface's  extremal cross-section normal to \textit{z}. The warped-sphere model, characteristic of III-V systems, makes the tracking and fitting of oscillations fairly straight forward. In this manner, if the angle between the surface normal of the sample and the magnetic field is scanned through a mirror symmetry plane one can deduce whether mirror symmetry is present based on the if the oscillations frequency is symmetric about certain angles. For example, in Fig. \ref{fig:SdH} we present a Monte Carlo simulation of a quantum oscillations experiment. A) is a general overview of device structure with the important parts being the 4-probe wiring and the scanning angle defined off the sample's  $[\bar110]$ direction. The model compliance tensor used is the same as Fig. \ref{fig:TPD} and the spin orbit coupling parameter $\gamma$ of 35 $eV \AA^3$ is approximated from InSb (without loss of generality, here C, D of 350 $eV \AA$, 35 $eV \AA$, respectively, are used). In this experimental configuration we can scan through the $(1\bar10)$ mirror plane as well as symmetric projections and record the oscillation frequencies as a function of angle. 
The results for two programmatically generated Fermi surfaces at 0.1 eV above the conduction band edge is then presented as C) for a full $2\pi$ rotation. The unstrained control sample is reasonably symmetric about $(1\bar10)$ \cite{seiler1977inversion}, whereas the strained sample has peaks of different heights, indicating that the $(1\bar10)$ mirror symmetry is broken.
The sensitivity of the Fermi surface on strain is apparent both in the band structure of Fig. \ref{fig:spin} and the inset of Fig. \ref{fig:SdH}.

\section{Conclusion}
We presented and discussed the prospects of strain engineering crystals lacking an inversion center into topological hosts of Kramers-Weyl node. The spin-texture, band structure and topological phase space for strained zinc blende is investigated and explained. Lastly, we presented the effect of strain on quantum oscillations and how to identify the desired symmetry breaking. While we kept the scope of the conversation limited to zinc blende materials which lack an inversion center; by symmetry considerations, alike behaviors of strain-enabled Kramers-Weyl phase can be extended to other systems lacking an inversion center\textemdash such as wurtzite and other piezoelectrics.  Even more generally, systems that possess an inversion center can be considered if we strain-engineer inversion asymmetry in addition to the mirror/rotoinversion symmetry-breaking geometric deformations discussed in this work.\\

\section{Acknowledgment}
We acknowledge the funding support from US AFOSR under award number FA9550-23-1-0310.

\nocite{*}

\bibliography{apssamp}
\end{document}